\documentclass[reprint,aps,prl,twocolumn,superscriptaddress,floatfix]{revtex4-1}
\usepackage{graphicx}
\usepackage{mathrsfs}
\usepackage{bm}
\usepackage{hyperref}
\hypersetup{hidelinks}
\usepackage{dcolumn}
\usepackage{amsmath}
\usepackage{amssymb}
\usepackage{bbold}
\usepackage{mathtools}
\usepackage{color}
\usepackage{array}

\usepackage[normalem]{ulem}

\begin{document}

\title{Nonlocal Einstein Relation in Confined Phonon Hydrodynamics}

\author{Guanda Quan}
\affiliation{School of Energy and Power Engineering, Huazhong University of Science and Technology, Wuhan 430074, China}
\author{Ronggui Yang}
\email{ronggui@pku.edu.cn}
\affiliation{School of Energy and Power Engineering, Huazhong University of Science and Technology, Wuhan 430074, China}
\affiliation{School of Mechanics and Engineering Science, Peking University, Beijing 100871, China}
\author{Xin Qian}
\email{xinqian21@hust.edu.cn}
\affiliation{School of Energy and Power Engineering, Huazhong University of Science and Technology, Wuhan 430074, China}

\begin{abstract}
Einstein relations connect carrier mobility to macroscopic transport coefficients and constitute a fundamental link between microscopic dynamics and transport. In phonon hydrodynamics, this connection is conventionally local: the drift induced thermal conductivity is proportional to the local drift mobility. Here we show that this relation becomes intrinsically nonlocal under spatial confinement. We derive a nonlocal Einstein relation, $\kappa_H(y)=\int \mathcal{P}(y-y')\mu(y')\,dy'$, in which hydrodynamic thermal conductivity $\kappa_H(y)$ at one position depends on collective drift mobility $\mu(y')$ across the entire cross-section of the channel. In the small-Knudsen-number limit, our nonlocal theory recovers the viscous heat equation and the local Einstein relation. At finite Knudsen number, nonlocality produces intrinsic boundary slip and distinct size-scaling behavior of the drift mobility $\mu$ and hydrodynamic thermal conductivity $\kappa_H$. First-principles calculations for graphite show that nonlocality broadens the channel-size window over which thermal conductivity exhibits superlinear size dependence. Temperature-dependent simulations further show that the nonlocal theory captures the emergence of a hydrodynamic peak in thermal conductivity normalized by ballistic conductance, whereas the local theory fails to accurately predict its magnitude. Our results establish a microscopic connection between collective phonon mobility and heat conduction beyond local hydrodynamics.
\end{abstract}

\maketitle

\textit{Introduction---} Einstein relations establish a fundamental connection between the mobility of microscopic carriers and macroscopic transport coefficients \cite{ref1,ref2}. In their conventional form, this connection is local: the transport response at a given position is determined by the carrier dynamics at the same position. This locality applies when microscopic correlations decay over length scales much shorter than the mean free paths of carriers. Hydrodynamic transport challenges this assumption because momentum is collectively redistributed over finite distances before relaxation \cite{ref3}. In a spatially confined system, the momentum correlation length can be comparable to the system dimensions, calling into question the locality of the Einstein relation itself.

Phonon hydrodynamics provides a particularly revealing setting for addressing this question. Beyond the diffusive and quasiballistic phonon kinetics \cite{ref4,ref5,ref6}, phonon hydrodynamics emerges when momentum-conserving normal (N) scattering dominates over momentum-relaxing resistive (R) scattering \cite{ref7}, resulting in a fluid-like behavior \cite{ref7,ref8,ref9,ref10,ref11}. At the continuum level, phonon hydrodynamics can be described as a viscous fluid \cite{ref12}, whose constitutive relations are local \cite{ref13,ref14}:

\begin{equation}
\label{eq:1}
J_q^i=-\kappa_K^{ij}\frac{\partial T}{\partial x^j}+\alpha^{ij}u^j,
\end{equation}
\begin{equation}
\label{eq:2}
J_P^{ij}=\beta^{ij}\frac{\partial T}{\partial x^j}-\eta^{ijkl}\frac{\partial u^k}{\partial x^l},
\end{equation}
where \(J_q\) and \(J_P\) are heat and momentum fluxes, \(\kappa_{K}\) is the kinetic thermal conductivity without phonon drifting, \(\alpha\) and \(\beta\) are energy-momentum coupling tensors, and \(\eta\) is the dynamic viscosity of phonons \cite{ref12}. In the linear transport regime \cite{ref15}, a drifting mobility \(\mu\) can be defined as \(u^{i}=-\mu^{ij}\partial T/\partial x^{j}\), and the heat flux can be written as \cite{ref16} \(J_q^i=-\left(\kappa_K^{ij}+\kappa_H^{ij}\right)\partial T/\partial x^j\), where \(\kappa_{H}^{ij}\) is the hydrodynamic thermal conductivity due to phonon drifting:

\begin{equation}
\label{eq:3}
\kappa_H^{ij}=\alpha^{ik}\mu^{kj}.
\end{equation}
Eq.~(\ref{eq:3}) is the Einstein relation for phonon hydrodynamics.

However, the locality underlying Eq.~(\ref{eq:3}) becomes questionable in confined hydrodynamic flow. Phonon Poiseuille flow occurs when \((\Lambda_{N} \ll W \ll \Lambda_{R})\), where \(W\) is the channel width and \(\Lambda_{N\ or\ R}\) is the characteristic mean free path of N or R scattering, respectively \cite{ref7}. While frequent N scattering redistributes phonon energy and establishes local equilibrium, it does not relax the collective momentum. Momentum correlations can therefore persist over a substantial fraction of the channel, allowing the boundary to influence transport far into the interior. This coexistence of local equilibration and long-ranged momentum transport raises a fundamental question: how is collective phonon mobility related to thermal conductivity when hydrodynamic transport becomes spatially nonlocal?

Here we show that the Einstein relation of phonon transport cease to be normal under spatial confinement. Instead of the local proportionality, the hydrodynamic thermal conductivity is related to the collective drift mobility through spatial convolution, while the drift itself is governed by a nonlocal Fredholm integral equation. Our theory recovers the VHE \cite{ref12} and the local Einstein relation at the small Knudsen number limit. At finite Knudsen numbers, spatial nonlocality leads to an intrinsic boundary slip that breaks the commonly assumed nonslip condition, a flattened hydrodynamic heat-flux profile, and a broadened hydrodynamic window with superlinear size scaling. Further, our nonlocal theory captures the emergence of the hydrodynamic peak in temperature dependent thermal transport, whereas the local theory fails to predict its magnitude. Our work establishes a fundamental connection between collective mobility and transport coefficients beyond local hydrodynamics.

\textit{The Nonlocal Theory---} We consider steady-state phonon Poiseuille flow across a channel under a constant temperature gradient \(-\nabla T=-\frac{dT}{dx}\cdot\widehat{\mathbf{x}}\) aligned with the \(x\)-axis. For clarity of derivation, we use the Peierls-Boltzmann transport equation (PBTE) under the Callaway scattering approximation (CSA) to obtain the nonlocal Einstein relation. We will show that using the full scattering matrix \cite{ref17,ref18} obtain identical results in Note S1 of Supplemental Material \cite{ref19} despite the complicated algebra. The steady-state PBTE under CSA is written as \cite{ref20,ref21,ref22}:

\begin{equation}
\label{eq:4}
\mathbf{v}_\lambda\cdot\nabla n_\lambda=-\frac{n_\lambda}{\tau_\lambda^R}-\frac{n_\lambda^D}{\tau_\lambda^N},
\end{equation}
where \(\lambda\) is the phonon mode index, \(\tau^{R}\) and \(\tau^{N}\) are the relaxation times of the R and N processes; \(n_{\lambda}=N_{\lambda}-N_{\lambda}^{0}(T)\) the population deviation from the Bose-Einstein distribution \(N_{\lambda}^{0}(T)=\left[\exp\left(\hbar\omega_{\lambda}/k_{B}T\right)-1\right]^{-1}\); \(n_{\lambda}^{D}=N_{\lambda}-N_{\lambda}^{D}(T)\) is the deviation from drifted equilibrium \(N_{\lambda}^{D}(T)=N_{\lambda}^{0}(T)+\frac{C_{\lambda}\overline{T}}{\hbar\omega_{\lambda}^{2}}q_{\lambda}^{x}u\). In the well-developed region, we can separate the population deviation into \(n_{\lambda}=\frac{C_{\lambda}}{\hbar\omega_{\lambda}}\Delta T(x)-\frac{C_{\lambda}}{\hbar\omega_{\lambda}}\Lambda_{\lambda}^{x}\frac{dT}{dx}+Y_{\lambda}(y)\), where \(C_{\lambda}=\frac{(\hbar\omega_{\lambda})^{2}}{k_{B}\overline{T}^{2}}N_{\lambda}^{0}(N_{\lambda}^{0}+1)\) is the specific heat of phonon mode \(\lambda\). Eq.~(5) can therefore be reduced to:

\begin{equation}
\label{eq:5}
\left(1+\Lambda_\lambda^y\frac{d}{dy}\right)Y_\lambda(y)=\eta_\lambda\frac{C_\lambda\overline{T}}{\hbar\omega_\lambda^2}q_\lambda^x\cdot u(y),
\end{equation}
where \(\eta_{\lambda}=\tau_{\lambda}/\tau_{\lambda}^{N}\) denotes the relative scattering strength of N processes, \(\Lambda_{\lambda}^{i}=v_{\lambda}^{i}\tau_{\lambda}\) is the mean free path along the \(i\)-th direction. The solution to Eq.~(5) is written as

\begin{equation}
\label{eq:6}
\begin{split}
Y_\lambda^{\pm}(y)&=Y_b e^{-(y\pm d)/\Lambda_\lambda^y}\\
&\quad\pm\eta_\lambda\frac{C_\lambda\overline{T}}{\hbar\omega_\lambda^2}q_\lambda^x
\int_{\mathbb{C}^{\pm}}\frac{1}{\Lambda_\lambda^y}e^{-(y-y')/\Lambda_\lambda^y}u(y')\,dy',
\end{split}
\end{equation}
where the superscript \(\pm\) indicate phonons propagating along the \(\pm y\) direction. \(Y_b=\frac{C_\lambda\Lambda_\lambda^x}{\hbar\omega_\lambda}\frac{dT}{dx}\) is the value at a thermalizing boundary \cite{ref23}. The integral ranges are \(\mathbb{C}^{+}=[-d,y]\) and \(\mathbb{C}^{-}=[y,d]\). Substituting Eq.~(6) to the momentum conservation requirement \(\sum_\lambda \hbar\mathbf{q}_\lambda n_\lambda^D/\tau_\lambda^N=0\), the nonlocal Fredholm integral equation can be derived:

\begin{equation}
\label{eq:7}
\begin{split}u(\widehat{y})&=\left(-\frac{dT}{dx}\right)f(\widehat{y})\\&\quad+\int_{-1}^{1}K(\widehat{y}-\widehat{y}')u(\widehat{y}')\,d\widehat{y}',\end{split}
\end{equation}
where \(\widehat{y}=y/d\) is the dimensionless coordinate, the source function \(f(\widehat{y})\) and the spatial kernel \(K(\widehat{y}-\widehat{y}')\) are defined in Supplemental Table SI \cite{ref19}.

Dividing \(-dT/dx\) on both sides of Eq.~(7), we obtain the integral equation for drifting mobility:

\begin{equation}
\label{eq:8}
\mu(\widehat{y})=f(\widehat{y})+\int_{-1}^{1}K(\widehat{y}-\widehat{y}')\mu(\widehat{y}')\,d\widehat{y}'.
\end{equation}

We solve Eq.~(\ref{eq:7}) or Eq.~(\ref{eq:8}) numerically using Fourier series expansion \cite{ref23}, see Supplementary Note S2 \cite{ref19} for details.

The heat flux can be calculated as \(J_{q} = \sum_{\mu}{\hbar\omega_{\mu}v_{\mu}^{x}n_{\mu}}\). Using the PBTE solution, we can derive the following equations:

\begin{equation}
\label{eq:9}
J_q=J_{qK}+J_{qH}=-\kappa_K(\widehat{y})\frac{dT}{dx}-\kappa_H(\widehat{y})\frac{dT}{dx},
\end{equation}
\begin{equation}
\label{eq:10}
\kappa_K(\widehat{y})=\sum_\lambda C_\lambda v_\lambda^x\Lambda_\lambda^x-\Delta\kappa_{\mathrm{surf}}(\widehat{y}),
\end{equation}
\begin{equation}
\label{eq:11}
\begin{split}\Delta\kappa_{\mathrm{surf}}(\widehat{y})&=\sum_{v_\lambda^y\geq0}C_\lambda v_\lambda^x\Lambda_\lambda^x e^{-(\widehat{y}+1)/|\mathrm{Kn}_\lambda^y|}\\&\quad+\sum_{v_\lambda^y<0}C_\lambda v_\lambda^x\Lambda_\lambda^x e^{-(1-\widehat{y})/|\mathrm{Kn}_\lambda^y|}.\end{split}
\end{equation}
\begin{equation}
\label{eq:12}
\kappa_H(\widehat{y})=\mathcal{P}*\mu=\int_{-1}^{1}\mathcal{P}(\widehat{y}-\widehat{y}')\mu(\widehat{y}')\,d\widehat{y}',
\end{equation}
where \(\mathrm{Kn}_{\lambda}^{y}=\Lambda_{\lambda}^{y}/d\) is the mode-resolved Knudsen number.

\begin{figure}[t]
\centering
\includegraphics[width=\columnwidth]{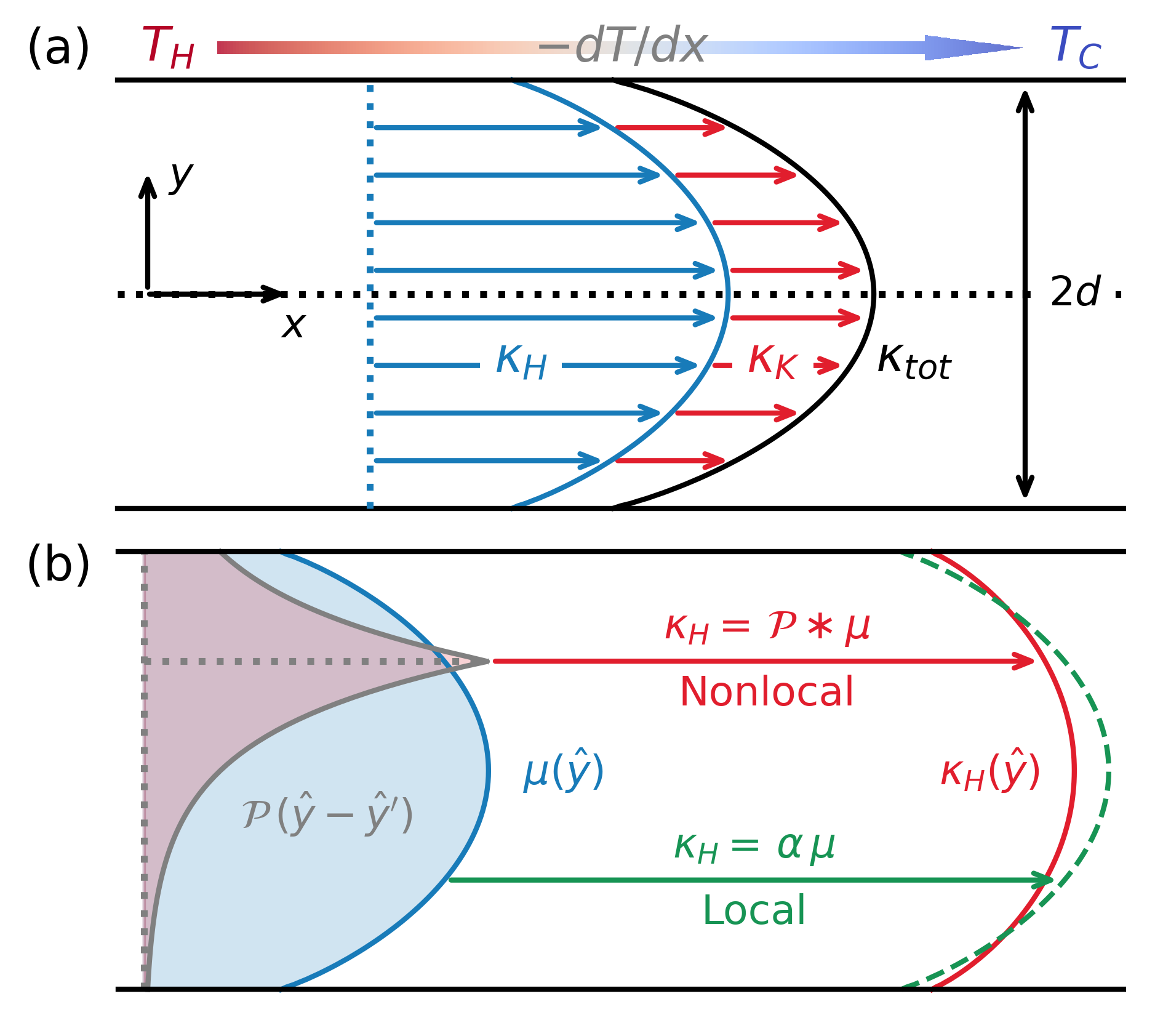}
\caption{(a) Schematic of steady-state phonon Poiseuille flow in a heat conduction channel. The thermal conductivity \(\kappa_{tot}(\widehat{y})\) contains a kinetic part \(\kappa_{K}(\widehat{y})\) and a hydrodynamic part \(\kappa_{H}(\widehat{y})\). (b) Illustration of the nonlocal and local Einstein relation for \(\kappa_{H}(\widehat{y})\). In the local Einstein relation, \(\kappa_{H}(\widehat{y})\) is determined by the mobility \(\mu(\widehat{y})\) at the same location, while the nonlocal Einstein relation says \(\kappa_{H}(\widehat{y})\) is determined by a spatial convolution \(\mathcal{P}*\mu\).}
\label{fig:overview}
\end{figure}

Eqs.~(\ref{eq:9})--(\ref{eq:12}) are the major analytical results of this work. The total heat flux can be separated into a kinetic component \(J_{qK}\) due to diffusive/quasiballilstic motion of phonons, and a hydrodynamic component \(J_{qH}\) due to collective drift. Consequently, the total thermal conductivity is also a summation of the kinetic conductivity \(\kappa_{K}\) and the hydrodynamic conductivity \(\kappa_{H}\), as shown in Fig.~\ref{fig:overview}(a). The kinetic conductivity is the bulk conductivity reduced by a surface term \(\Delta\kappa_{surf}\), previously described as boundary friction by Cepellotti and Marzari \cite{ref24}. The hydrodynamic thermal conductivity \(\kappa_{H}\) is defined as the nonlocal Einstein relation in Eq.~(\ref{eq:12}), with \(*\) denoting the convolution product and the transport kernel \(\mathcal{P}\) defined in Supplemental Table SI \cite{ref19}. Eq.~(\ref{eq:12}) is schematically illustrated in Fig.~\ref{fig:overview}(b). The local hydrodynamic conductivity \(\kappa_{H}(\widehat{y})\) at \(\widehat{y}=y/d\) is an integral result of the hydrodynamic drift across the channel. At the local limit (\(\left|\mathrm{Kn}_{\lambda}^{y}\right|\ll1\)), \(\mathcal{P}(\widehat{y}-\widehat{y}')\rightarrow\alpha\delta(\widehat{y}-\widehat{y}')\), and Eq.~(\ref{eq:12}) is reduced to the local Einstein relation \(\kappa_{H}=\alpha\mu\), with the temperature-momentum coupling coefficient defined as \(\alpha=\overline{T}\sum_{\lambda}\eta_{\lambda}C_{\lambda}q_{\lambda}^{x}v_{\lambda}^{x}/\omega_{\lambda}\).

While we have used CSA in the derivation, in the Supplemental Note S1 \cite{ref19}, we prove that the identical Fredholm integral equation and the nonlocal Einstein relation can still be obtained using the full scattering matrix (\emph{i.e.} the relaxon theory \cite{ref18}) ; hence the simplified scattering matrix does not affect the nonlocal transport physics in confined phonon hydrodynamics.

\textit{Convergence to VHE at the local limit---} In the local limit of \(\left|\mathrm{Kn}_{\lambda}^{y}\right|\ll1\), by expanding the Fredholm kernel and performing the convolution \(K*u\), the following differential equation can be obtained to the leading order:

\begin{equation}
\label{eq:13}
\beta\frac{\partial T}{\partial x}+\eta\frac{d^2u}{dy^2}=-\gamma u,
\end{equation}
where \(\gamma=\sum_{\lambda}\frac{\eta_{\lambda}C_{\lambda}(q_{\lambda}^{x})^{2}}{\omega_{\lambda}^{2}\tau_{\lambda}^{R}}\) is the damping coefficient of collective drift, \(\beta=\alpha/\overline{T}\) is the momentum-temperature coupling coefficient, and \(\eta=\sum_{\lambda}\frac{\eta_{\lambda}C_{\lambda}(q_{\lambda}^{x}\Lambda_{\lambda}^{y})^{2}}{\omega_{\lambda}^{2}\tau_{\lambda}^{N}}\) is the dynamic viscosity of phonons. Eq.~(\ref{eq:13}) is identical to VHE at the one-dimensional steady state limit. Detailed derivation of Eq.~(\ref{eq:13}) is shown in Supplemental Note S3 \cite{ref19}. Under the nonslip boundary condition, the drifting velocity profile is

\begin{equation}
\label{eq:14}
u_{\mathrm{local}}(y)=u_{\mathrm{bulk}}\left[1-\frac{\cosh(y/\ell)}{\cosh(d/\ell)}\right],
\end{equation}
where \(u_{bulk}=-\frac{\beta}{\gamma}\frac{dT}{dx}\) is the drifting velocity in the bulk limit, and \(\ell^{2}=\eta/\gamma\) is the hydrodynamic length. The parabolic Poiseuille profile is further recovered when \(d\ll\ell\):

\begin{equation}
\label{eq:15}
u_{\mathrm{local}}(y)=-\frac{\beta}{2\eta\overline{T}}\frac{dT}{dx}(d^2-y^2).
\end{equation}

\textit{Breakdown of Nonslip Boundary Condition---} The first nonlocal effect is the breakdown of the nonslip boundary condition for the drifting velocity. By performing a boundary expansion of the kernel \(K_{\lambda}\), we can derive the boundary slip as:

\begin{equation}
\label{eq:16}
u_w=\overline{\mathrm{Kn}}\left.\left(\frac{\partial u}{\partial\widehat{y}}\right)\right|_{\widehat{y}=\pm1}.
\end{equation}
See detailed derivation in Supplemental Note S4 \cite{ref19}. \(\overline{Kn}\) is the average Knudsen number:

\begin{equation}
\label{eq:17}
\overline{\mathrm{Kn}}=\frac{\sum_\lambda \dfrac{\eta_\lambda C_\lambda(q_\lambda^x)^2}{\omega_\lambda^2\tau_\lambda^N}|\mathrm{Kn}_\lambda^y|}{\sum_\lambda \dfrac{\eta_\lambda C_\lambda(q_\lambda^x)^2}{\omega_\lambda^2\tau_\lambda}}.
\end{equation}

\begin{figure}[t]
\centering
\includegraphics[width=\columnwidth]{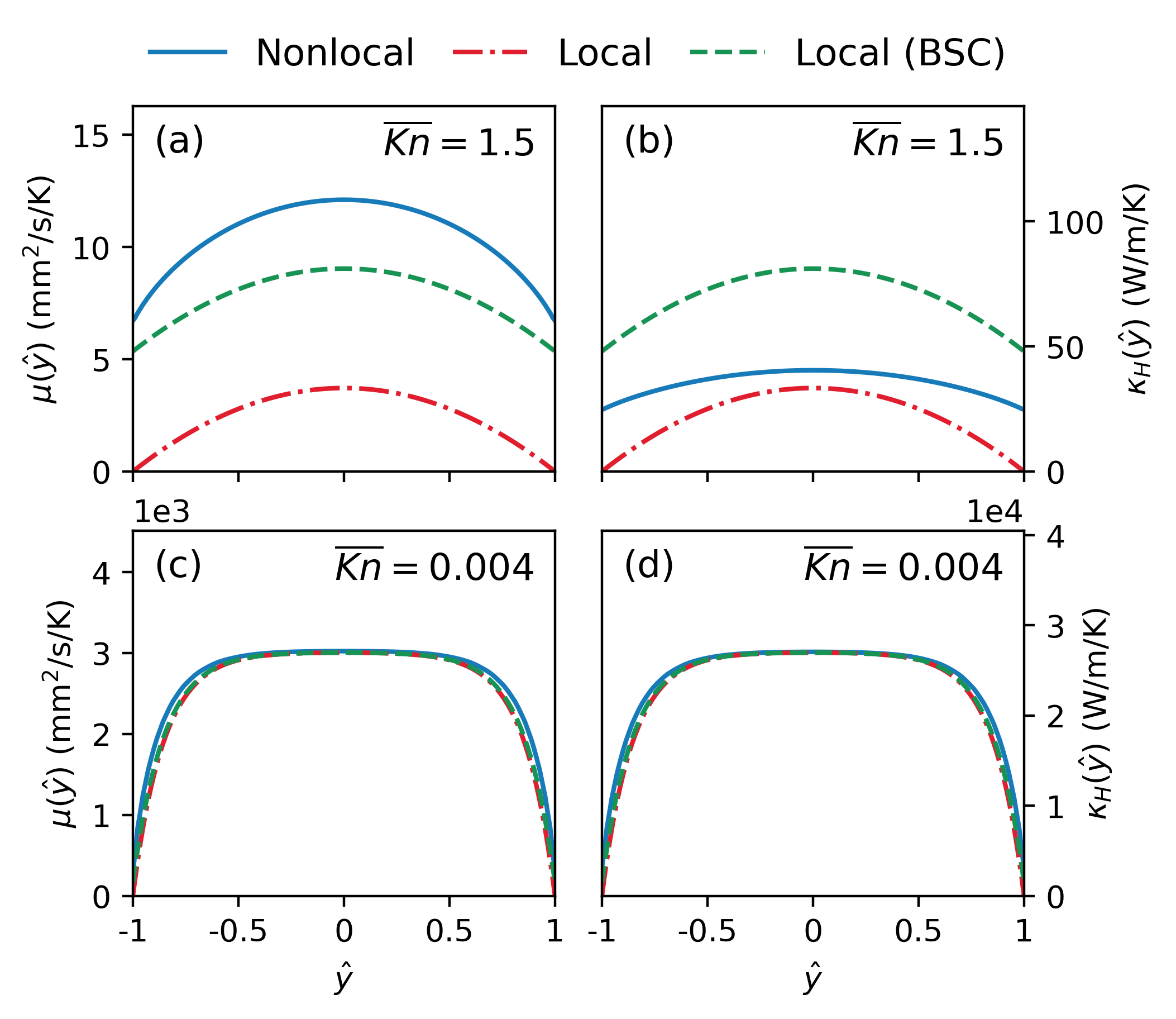}
\caption{Cross-sectional profiles of (a) drifting mobility \(\mu(\widehat{y})\) and (b) hydrodynamic thermal conductivity \(\kappa_{H}(\widehat{y})\) at an average Knudsen number of 1.5. Drifting mobility (c) and (d) hydrodynamic thermal conductivity profiles at an averaged Knudsen number of 0.004. Local (BSC) refers to simulation using local VHE while using a boundary slip correction.}
\label{fig:profiles}
\end{figure}

Fig.~\ref{fig:profiles} shows the numerical results of a graphite strip from \emph{ab initio} phonon properties (see details in Supplemental Note S5 \cite{ref19}), compared with the local model (Eq.~(\ref{eq:13})). The nonlocal Fredholm equation results in a finite drift velocity \(u_{w}\) and therefore a finite \(\kappa_{H}\) at the boundary. Only when \(\overline{Kn} \ll 1\), the drifting velocity at the boundary approaches zero. Another nonlocal consequence is that \(\kappa_{H}(\widehat{y})\) becomes flattened compared with \(\mu(\widehat{y})\), because \(\mathcal{P}(y-y')\) serves as a weighed spatial average across the channel width.

\begin{figure}[t]
\centering
\includegraphics[width=\columnwidth]{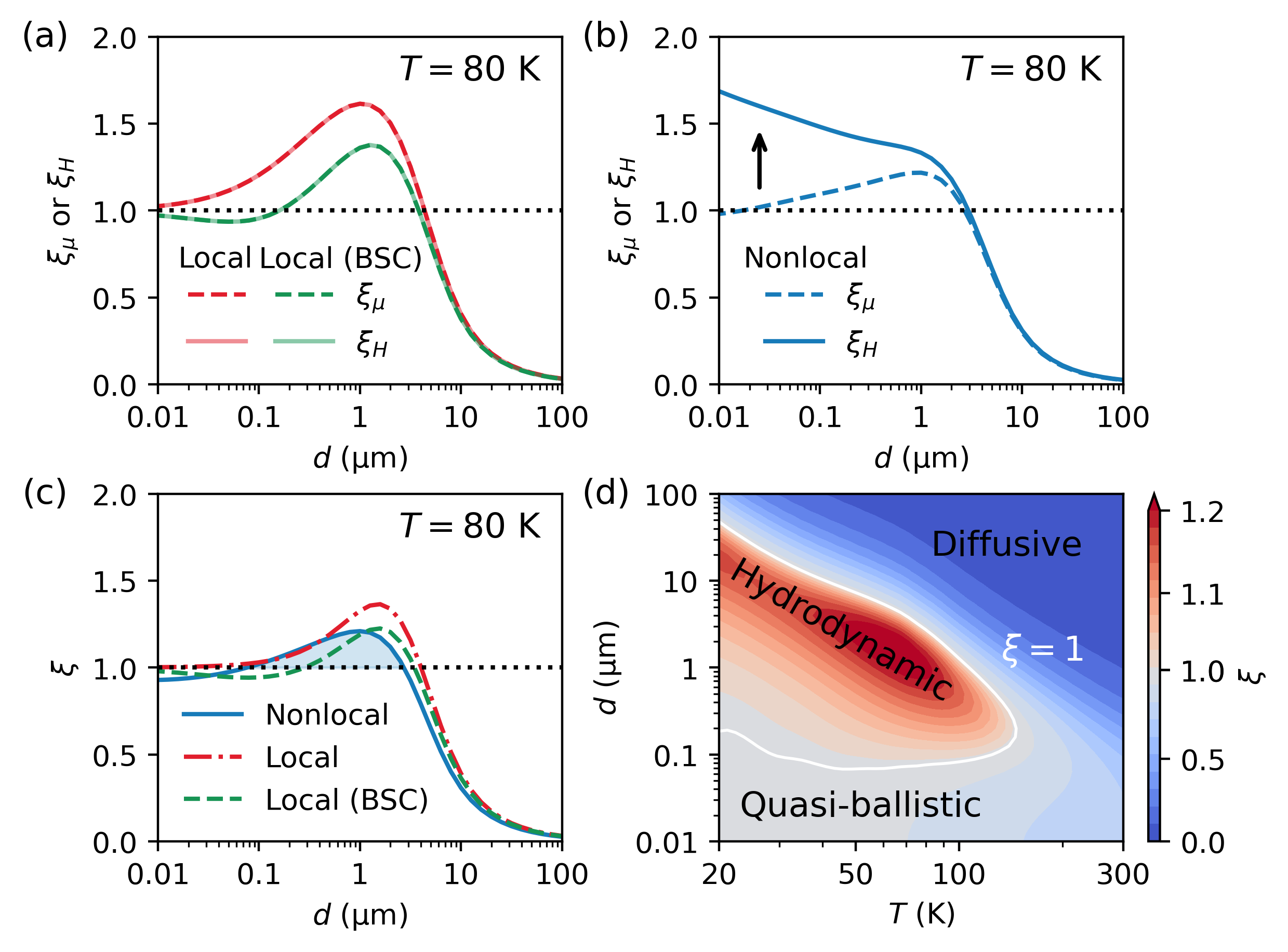}
\caption{The scaling exponent for effective mobility and effective hydrodynamic thermal conductivity in the (a) nonlocal model (this work) and (b) local models with/without boundary slip at 80 K. The arrow indicates the difference in the scaling exponent caused by the nonlocal Einstein relation. (c) Scaling exponents of effective thermal conductivity in different models at 80 K and the shaded hydrodynamic regime for the nonlocal model; (d) Scaling exponent of the nonlocal model at different temperatures and sizes, with contour lines for \(\xi = 1\) plotted to delineate the hydrodynamic, quasi-ballistic, and diffusive regions.}
\label{fig:scaling}
\end{figure}

\textit{Channel Size Scaling---} The nonlocal Einstein relation further results in decoupled size scaling of \(\kappa_{H}\) and \(\mu\). In Fig.~\ref{fig:scaling}, we show the scaling exponents of \(\langle\kappa_{H}\rangle\sim d^{\xi_{H}}\) and \(\langle\mu\rangle\sim d^{\xi_{\mu}}\) at different channel size \(d\), where \(\left\langle f \right\rangle = \frac{1}{2d}\int_{-d}^{d}f(y)\,dy\) denotes the cross-sectional average. As shown in Fig.~\ref{fig:scaling}(a), the local Einstein relation \(\kappa_{H}=\alpha\mu\) ensures identical size scaling of \(\kappa_{H}\) and \(\mu\), i.e. \(\xi_{H}=\xi_{\mu}\). If the local theory is corrected by the boundary slip (Eq.~(\ref{eq:16})), the superlinear scaling exponent becomes less pronounced, because the boundary slip contributes to a linear scaling component upon the quadratic Poiseuille profile. Further, the scaling identity \(\xi_{H}=\xi_{\mu}\) is broken if the nonlocal Einstein relation is used to compute \(\kappa_{H}\) (Fig.~\ref{fig:scaling}(b)). As channel size reduces, the drifting mobility recovers the ballistic linear scaling, but the hydrodynamic \(\kappa_{H}(d)\) can still reduce superlinearly. In the nonlocal picture, the collective drift momentum approaching the boundary can get dissipated before it reaches the exact boundary wall, which enhances the size dependence compared with the local model. In Fig.~\ref{fig:scaling}(c), we calculated the scaling of total thermal conductivity \(\left\langle\kappa_{tot}\right\rangle=\left\langle\kappa_{K}\right\rangle+\left\langle\kappa_{H}\right\rangle\sim d^{\xi}\). The hydrodynamic size window is shaded with \(\xi\geq1\). The nonlocal effects contribute to a broadened hydrodynamic window of channel size, because the nonlocal effects facilitate collective phonon drift at small sample sizes. In Fig.~\ref{fig:scaling}(d), we calculated the scaling factor \(\xi\) at different temperatures, demonstrating that our nonlocal theory captures the transition from the quasiballistic regime \((0<\xi\leq1)\), to the superlinear hydrodynamic regime \((\xi\geq1)\) and the diffusive regime \((\xi\approx0)\).

\begin{figure}[t]
\centering
\includegraphics[width=\columnwidth]{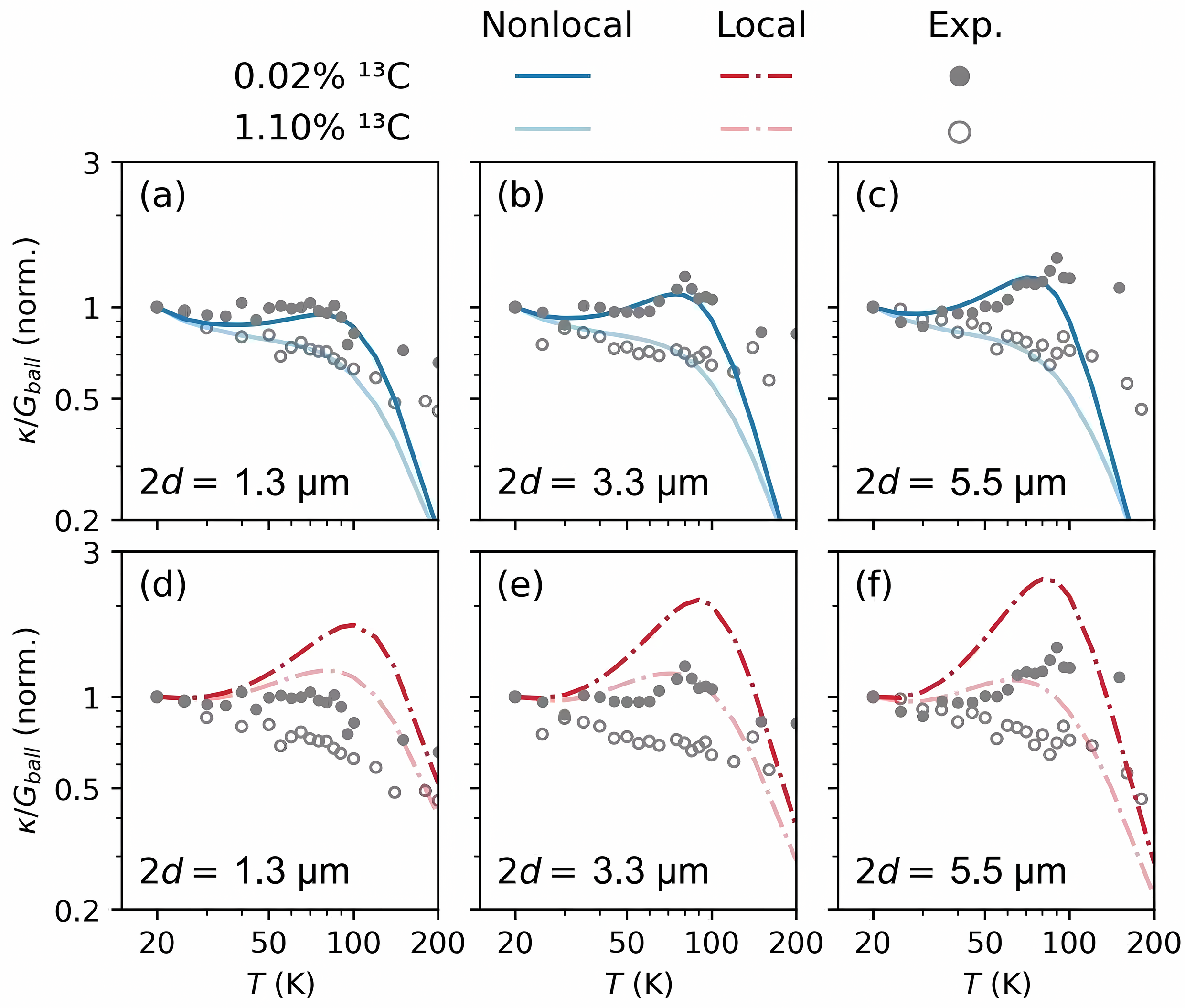}
\caption{Nonlocal theoretical predictions of \(\frac{\kappa}{G_{ball}}\) for natural and isotope-enriched graphite ribbons with widths of (a) 1.3 \(\mu\mathrm{m}\) (b) 3.3 \(\mu\mathrm{m}\) and (c) 5.5 \(\mu\mathrm{m}\). The local theory predictions are presented in (d-f) with the same channel widths in (a-c), respectively. Experimental data are taken from ref. \cite{ref25}.}
\label{fig:temperature}
\end{figure}

\textit{Temperature dependence---} While channel-size scaling provides a direct theoretical signature of nonlocal hydrodynamics, phonon Poiseuille flow is more commonly identified experimentally from the temperature dependence of \(\kappa/G_{ball}\), where \(G_{ball}\) is the ballistic thermal conductance \cite{ref25}. At low temperatures, ballistic transport gives only weak temperature dependence. As normal scattering strengthens within the hydrodynamic size window, collective phonon drift enhances heat transport and increases \(\kappa/G_{ball}\). At higher temperatures, resistive scattering dominates, leading to decreases in \(\kappa/G_{ball}\). The resulting crossover produces a characteristic peak in \(\kappa/G_{ball}\), fingerprinting the emergence of phonon hydrodynamics. Figure~\ref{fig:temperature} compares this temperature dependence with measurements on graphite ribbons of different widths and isotope concentrations. To normalize the absolute value difference caused by the experimental finite thickness, \(\kappa/G_{ball}\) is normalized by its value at 20 K. For isotope-enriched graphite, the nonlocal theory captures the emergence and growth of the hydrodynamic peak as the ribbon width increases from 1.3 to 5.5 \(\mu\mathrm{m}\), while isotope scattering suppresses the peak in natural graphite. The local theory systematically overestimates the hydrodynamic enhancement because it neglects the finite spatial range of boundary-induced momentum dissipation. As the width increases and the Knudsen number decreases, the nonlocal result converges to the local limit, as shown for a 20-\(\mu\mathrm{m}\) ribbon in Supplemental Fig. S2. These results demonstrate that the nonlocal theory consistently describes the ballistic--hydrodynamic--diffusive crossover with increasing temperature under confinement.

\textit{Conclusion---} We have established a nonlocal Einstein relation for confined phonon hydrodynamics, in which the hydrodynamic thermal conductivity is related to collective phonon mobility through a spatial convolution rather than a local proportionality. Together with the Fredholm equation governing collective drift, this framework provides a microscopic description of phonon hydrodynamics across the quasiballistic-to-diffusive crossover and recovers the viscous heat equation and the local Einstein relation in the small-Knudsen-number limit. At finite Knudsen number, spatial nonlocality gives rise to intrinsic boundary slip and decouples the spatial profiles and size scaling of collective mobility and hydrodynamic thermal conductivity. First-principles calculations for graphite further show that these nonlocal effects modify both the size dependence and the temperature-dependent hydrodynamic enhancement of thermal conductivity, providing experimentally accessible signatures beyond local hydrodynamic theory. Our results establish spatial nonlocality as an intrinsic feature of confined hydrodynamic transport and extend the connection between mobility and transport coefficients beyond the local regime. More broadly, the nonlocal Einstein relation developed here may provide a general framework for hydrodynamic transport of other collective carriers, such as electrons \cite{ref26} and magnons \cite{ref27}.

\textit{Acknowledgment---} X.Q. acknowledges funding support from the National Key R\&D Program (No. 2022YFA1203100) and the National Natural Science Foundation (NSFC No. U25D9006). R.Y. acknowledges support from NSFC under the Grant No. T2588301. The authors declare no conflict of interest.

\end{document}